\documentclass[a4paper, conference]{IEEEtran}
\usepackage[T1]{fontenc}
\usepackage[utf8]{inputenc}
\usepackage{cite}
\usepackage{amsmath,amssymb,amsfonts}
\usepackage{algorithmic}
\usepackage{graphicx}
\usepackage{textcomp}
\usepackage{xcolor}
\usepackage{subcaption}
\usepackage{tikz}
\usetikzlibrary{
    positioning,     
    arrows.meta,     
    shapes.geometric 
}
\usepackage{eso-pic}
\usepackage{multirow} 
\usepackage{booktabs} 
\usepackage{standalone}
\usepackage{siunitx}
\usepackage{stfloats}
\def\BibTeX{{\rm B\kern-.05em{\sc i\kern-.025em b}\kern-.08em
    T\kern-.1667em\lower.7ex\hbox{E}\kern-.125emX}}
\usepackage[acronym]{glossaries}

\newacronym{NVC}{NVC}{Neural Video Codec}
\newacronym{VCT}{VCT}{Video Compression Transformer}
\newacronym{RD}{RD}{Rate-Distortion}
\newacronym{CNN}{CNN}{Convolutional Neural Network}
\newacronym{MLP}{MLP}{Multi-Layer Perceptron}
\newacronym{BD}{BD}{Bjøntegaard Delta}
\newacronym{LRP}{LRP}{Latent Residual Prediction}
\newacronym{SWA}{SWA}{Sliding Window Attention}
\begin{document}
\bstctlcite{BSTcontrol}
\IEEEoverridecommandlockouts

\AddToShipoutPictureBG*{%
  \AtPageUpperLeft{%
    \raisebox{-3\baselineskip}{
      \makebox[\paperwidth][c]{
        \parbox{0.9\textwidth}{\centering\footnotesize 
          © 2025 IEEE. Personal use of this material is permitted. Permission from IEEE must be obtained for all other uses, in any current or future media, including reprinting/republishing this material for advertising or promotional purposes, creating new collective works, for resale or redistribution to servers or lists, or reuse of any copyrighted component of this work in other works.
        }%
      }%
    }%
  }%
}

\title{Sliding Window Attention for Learned Video Compression}

\author{\IEEEauthorblockN{Alexander Kopte and Andr\'{e} Kaup}
\IEEEauthorblockA{\textit{Multimedia Communications and Signal Processing} \\
\textit{Friedrich-Alexander University Erlangen-Nuremberg}\\
Erlangen, Germany \\
\{alex.kopte, andre.kaup\}@fau.de}}

\maketitle

\begin{abstract}
To manage the complexity of transformers in video compression, local attention mechanisms are a practical necessity. The common approach of partitioning frames into patches, however, creates architectural flaws like irregular receptive fields. When adapted for temporal autoregressive models, this paradigm, exemplified by the \gls{VCT}, also necessitates computationally redundant overlapping windows. This work introduces 3D \gls{SWA}, a patchless form of local attention. By enabling a decoder-only architecture that unifies spatial and temporal context processing, and by providing a uniform receptive field, our method significantly improves rate-distortion performance, achieving Bjøntegaard Delta-rate savings of up to \SI{18.6}{\percent} against the \gls{VCT} baseline. Simultaneously, by eliminating the need for overlapping windows, our method reduces overall decoder complexity by a factor of \num{2.8}, while its entropy model is nearly \num{3.5} times more efficient. We further analyze our model's behavior and show that while it benefits from long-range temporal context, excessive context can degrade performance.
\end{abstract}


\glsresetall

\section{Introduction}

Learned video compression is an active research field, with recent \glspl{NVC} beginning to outperform traditional codecs \cite{li_neural_2023, chen_channel-wise_2023}. While transformers have become dominant in many computer vision domains, most state-of-the-art \glspl{NVC} still rely on \glspl{CNN}. A common strategy to adapt transformers for video is to partition frames into non-overlapping patches, a paradigm used by \gls{VCT} \cite{mentzer_vct_2022} and Swin-based models \cite{zou_devil_2022, liu_learned_2023, zhu_transformer-based_2021, xiang_mimt_2022}. However, this approach introduces a key architectural flaw: the receptive field for each hyperpixel is defined relative to its enclosing patch instead of the hyperpixel itself, making the local context position-dependent. This issue is particularly detrimental in a strictly autoregressive setting, where causal masking results in an unequal number of visible neighbors for different hyperpixels within the same patch.

The patch-based approach in our primary baseline, \gls{VCT}, also introduces hard boundaries that impede information flow between patches. This issue becomes particularly significant for temporal context, as motion can cause features corresponding to a patch in the current frame to lie in entirely different patches in previous frames. To counteract this, \gls{VCT} processes larger, overlapping windows for temporal context extraction. This strategy, however, creates significant computational overhead by repeatedly processing the same hyperpixels. Furthermore, its bi-directional temporal encoder is not autoregressive, preventing an efficient causal decoding process.

To address these limitations, we propose a decoder-only, fully autoregressive entropy model that uses 3D \gls{SWA} to jointly process spatial and temporal context. This unified architecture eliminates artificial boundaries and avoids the redundant computations of \gls{VCT}'s overlapping window approach, while providing a uniform receptive field. The proposed model yields \gls{BD}-rate savings of up to \SI{18.6}{\percent} over the \gls{VCT} baseline, while reducing total decoder kMACs by a factor of \num{2.8}, with the entropy model becoming nearly \num{3.5} times more efficient.

\section{Related Work}

While \glspl{CNN} have been standard for learned video compression, recent work increasingly integrates transformers to overcome the limitations of static filters. Transformers are used, for example, to replace the main feature transform with a Swin-based architecture \cite{zou_devil_2022}, or in hybrid models that run Swin blocks alongside convolutions for both the feature transform and the entropy model \cite{liu_learned_2023}. Beyond the feature transform, transformers are also leveraged for other codec components, such as motion compression \cite{zhu_transformer-based_2021} or as entropy models \cite{xiang_mimt_2022}. The transformer-based entropy model in our primary baseline, \gls{VCT} \cite{mentzer_vct_2022}, uses a non-autoregressive, bi-directional temporal encoder on overlapping patches from two reference frames, and an autoregressive decoder that fuses this context with spatial information from the current frame's non-overlapping patches.


In contrast, our work moves away from patching entirely, instead building on local or sliding window attention. This mechanism has been successfully applied to 2D vision tasks as Neighborhood Attention \cite{hassani_neighborhood_2023}. The work of Chen \textit{et al.} \cite{chen_channel-wise_2023} also uses a non-patch-based transformer, but its application within the codec is fundamentally different from ours. They employ channel-wise attention as part of the feature transform and context fusion. In contrast, our model uses spatio-temporal attention exclusively within the entropy model to autoregressively predict the probability distribution of the quantized latents. This paper introduces the first fully spatio-temporal, non-patch-based, and fully autoregressive transformer as an entropy model for learned video compression.

\section{Proposed Method}

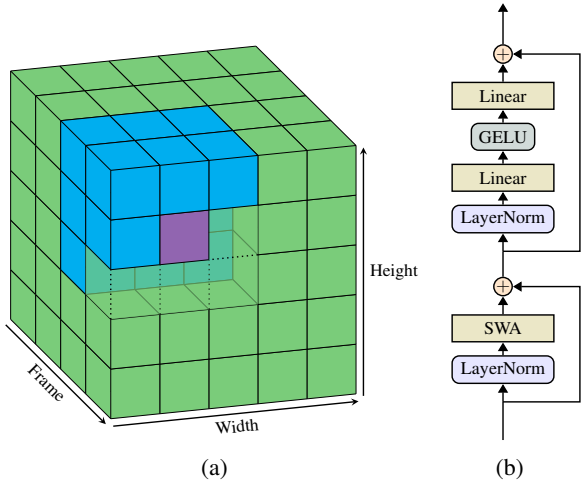
\begin{figure}[!t]
    \centering
    \begin{subfigure}[b]{0.31\textwidth}
        \centering
        \includestandalone[width=0.98\textwidth]{sliding_window}
        \caption{}
        \label{proposed:sliding:kernel}
    \end{subfigure}
    \begin{subfigure}[b]{0.105\textwidth}
        \centering
            \definecolor{greenish}{RGB}{235, 231, 199}
    \definecolor{teelish}{RGB}{210, 219, 217}
	\begin{tikzpicture}[
            attention/.style={
                rectangle, 
                draw, 
                thick, 
                text width=1.75cm,
                minimum height=1.5em,
                fill=greenish,
                align=center
            },
            projection/.style={
                rectangle, 
                draw, 
                thick, 
                text width=1.75cm,
                minimum height=1.5em,
                fill=greenish,
                align=center
            },
            activation/.style={
                rectangle, 
                draw, 
                thick, 
                text width=1cm,
                minimum height=1.5em,
                fill=teelish,
                align=center,
                rounded corners
            },
            norm/.style={
                rectangle, 
                draw, 
                thick, 
                text width=1.75cm,
                minimum height=1.5em,
                fill=blue!10,
                align=center,
                rounded corners
            },
            add/.style={
                circle, 
                draw, 
                thick, 
                fill=orange!20,
                inner sep=0pt,
                minimum size=3mm
            },
            line/.style={
                draw, 
                thick, 
                -Triangle 
            },
            path/.style={
                draw, 
                thick, 
                - 
            }
        ]

    \coordinate (input);
    \path[path] (input) -- ++(0, 0.75) coordinate(split1);

    \node[norm, above=0.35cm of split1] (norm1) {LayerNorm};
    \node[attention, above=0.25cm of norm1] (swa1) {SWA};
    \node[add, above=0.35cm of swa1] (add1) {$+$};

    \path[line] (split1) -- ++(1.5, 0) coordinate(resr11) -- (add1 -| resr11) coordinate(resr12) -- (add1);
	\path[line] (split1) -- (norm1);
    \path[line] (norm1) -- (swa1);
    \path[line] (swa1) -- (add1);

    \coordinate[above=0.5cm of add1] (split2);
    \path[path] (add1) -- (split2);
    
	\node[norm, above=0.35cm of split2] (norm2) {LayerNorm};
    \node[projection, above=0.25cm of norm2] (up1) {Linear};
    \node[activation, above=0.25cm of up1] (gelu1) {GELU};
    \node[projection, above=0.25cm of gelu1] (down1) {Linear};
    \node[add, above=0.35cm of down1] (add2) {$+$};

    \path[line] (split2) -- ++(1.5, 0) coordinate(resr21) -- (add2 -| resr21) coordinate(resr22) -- (add2);
	\path[line] (split2) -- (norm2);
    \path[line] (norm2) -- (up1);
    \path[line] (up1) -- (gelu1);
    \path[line] (gelu1) -- (down1);
    \path[line] (down1) -- (add2);
    \path[line] (add2) -- ++(0, 1) coordinate(output);
		
	\end{tikzpicture}
	
        \caption{}
        \label{proposed:sliding:architecture}
    \end{subfigure}
    \caption{Visualization of the proposed method. (a) The 3D \gls{SWA} kernel. For the current hyperpixel (purple), attention is computed over its previously decoded spatio-temporal neighbors (blue). Future hyperpixels that fall within the window (transparent) are causally masked. (b) The proposed transformer block architecture.}
    \label{proposed:sliding}
\end{figure}

\subsection{Sliding Window Attention}

Our proposed attention mechanism operates on the latent volume $y \in \mathbb{R}^{L \times H \times W \times C}$, where $L$, $H$, $W$, $C$ are the number of frames, height, width, and channels, respectively. 
The mechanism is analogous to a 3D convolution, applying a local kernel to each hyperpixel to gather context, as visualized in Fig. \ref{proposed:sliding:kernel}. For a given hyperpixel, attention is computed over its spatio-temporal neighbors within the window. Causal masking is applied to ensure that only previously decoded hyperpixels are used, while future ones are ignored. Compared to a standard convolution, our attention kernel has two key differences. First, the weights are not static but are dynamically computed based on the attention scores. Second, no zero-padding is applied at the boundaries of the latent volume, as this would be ill-suited for attention mechanisms due to introducing uninformative tokens into the context. Instead, the attention window is simply truncated, meaning the receptive field becomes progressively smaller for hyperpixels closer to an edge. However, since this effect is confined to the perimeter of the latent volume, its overall impact is minimal.

\begin{figure}[!t]
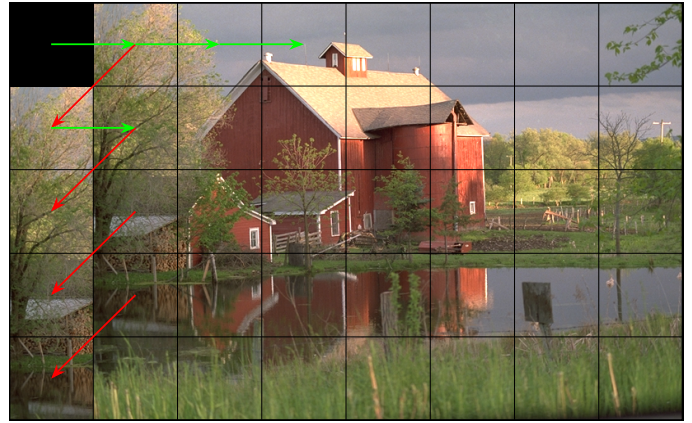

    \centering
    \includestandalone[width=0.489\textwidth]{prediction}
    \caption{Autoregressive prediction scheme for a single frame. Decoding follows a line-scan order (green). At the start of each new row, the hyperpixel directly above it is prepended to the input sequence (red), ensuring the local attention window has access to spatially relevant context.}
    \label{proposed:prediction}
\end{figure}

Formally, after flattening $y$ into a sequence $Y \in \mathbb{R}^{(L \cdot H \cdot W) \times C}$, we follow the standard multi-head attention paradigm \cite{vaswani_attention_2017}. First, for each of the $h$ attention heads, query ($Q_i$), key ($K_i$), and value ($V_i$) tensors are computed via linear projections with learnable weight matrices $W_i^Q, W_i^K, W_i^V$:
\begin{align}
    Q_i &= YW_i^Q, \quad K_i = YW_i^K, \quad V_i = YW_i^V 
\end{align}
The attention output for a single head is then calculated as:
\begin{align}
    A_i &= \text{Softmax}\left(\frac{Q_iK_i^T + B_i}{\sqrt{d_k}}\right)V_i 
\end{align}
where the scores are scaled by the square root of $d_k$, the dimension of the query and key vectors. Finally, the outputs of all $h$ attention heads are concatenated and projected with a final weight matrix $W^O$:
\begin{align}
    O &= \text{Concat}(A_1, \dots,A_h)W^O
\end{align}
Our key modification lies in the bias matrix $B_i \in \mathbb{R}^{(LHW)\times(LHW)}$, which enforces the local window. Its elements $b_i^{m,n}$ between hyperpixels $m$ and $n$ are defined by their relative positions $(\Delta l, \Delta y, \Delta x)$:
\begin{align}
    b_i^{m,n} = \begin{cases} 
        s_i^{\Delta l, \Delta y, \Delta x} & \text{if } |\Delta l| \le L_w, |\Delta y| \le H_w, |\Delta x| \le W_w \\
        -\infty & \text{otherwise}
    \end{cases}
\end{align}
Here, $s_i$ is a learnable tensor of size $(2L_w+1) \times (2H_w+1) \times (2W_w+1)$ containing the relative positional biases $s_i^{\Delta l, \Delta y, \Delta x}$ for one head, and $(L_w, H_w, W_w)$ defines the maximum offset from the center of the window along each axis. For efficient implementation, we adapt the FlashAttention-2 algorithm \cite{dao_flashattention-2_2023} in Triton \cite{triton} to avoid computing attention for fully masked blocks and to directly incorporate the biases from $s_i$.

\subsection{Entropy Model}

\begin{figure*}[!t]
    \centering
    \begin{minipage}{\textwidth}
        \centering
        \captionof{table}{\gls{BD}-rate savings (\%) relative to the \gls{VCT} baseline for all test datasets.}
        \label{experiments:results:table}
        \small
        \sisetup{
            retain-explicit-plus, 
            table-format = +-2.2 
          }
        \begin{tabular}{l l S S S}
            \toprule
            \textbf{Dataset} & \textbf{Coder} & \textbf{BD-Rate (I-frames)} & \textbf{BD-Rate (P-frames)} & \textbf{BD-Rate (GOP)} \\
            \midrule
            \multirow{5}{*}{UVG} & VCT & 0.0 \si{\percent} & 0.0 \si{\percent} & 0.0 \si{\percent} \\
            & SWA (ours) & -22.0 \si{\percent} & -17.2 \si{\percent} & -17.7 \si{\percent} \\
            & DCVC-DC & -31.0 \si{\percent} & -58.6 \si{\percent} & -55.1 \si{\percent} \\
            & HM 18.0 & +6.6 \si{\percent} & -19.3 \si{\percent} & -11.3 \si{\percent} \\
            & VTM 23.10 & -17.7 \si{\percent} & -43.8 \si{\percent} & -36.6 \si{\percent} \\
            \midrule
            \multirow{5}{*}{MCL-JCV} & VCT & 0.0 \si{\percent} & 0.0 \si{\percent} & 0.0 \si{\percent} \\
            & SWA (ours) & -22.6 \si{\percent} & -15.6 \si{\percent} & -16.2 \si{\percent} \\
            & DCVC-DC & -31.6 \si{\percent} & -50.7 \si{\percent} & -48.2 \si{\percent} \\
            & HM 18.0 & +6.4 \si{\percent} & -7.0 \si{\percent} & -1.8 \si{\percent} \\
            & VTM 23.10 & -17.3 \si{\percent} & -36.4 \si{\percent} & -31.3 \si{\percent} \\
            \midrule
            \multirow{5}{*}{HEVC B} & VCT & 0.0 \si{\percent} & 0.0 \si{\percent} & 0.0 \si{\percent} \\
            & SWA (ours) & -18.2 \si{\percent} & -18.6 \si{\percent} & -18.6 \si{\percent} \\
            & DCVC-DC & -29.1 \si{\percent} & -59.7 \si{\percent} & -56.0 \si{\percent} \\
            & HM 18.0 & +2.8 \si{\percent} & -21.9 \si{\percent} & -14.4 \si{\percent} \\
            & VTM 23.10 & -19.9 \si{\percent} & -49.7 \si{\percent} & -42.2 \si{\percent} \\
            \bottomrule
        \end{tabular}
    \end{minipage}
    \\[6pt] 

    \begin{subfigure}[b]{0.325\textwidth}
        \centering
        \includegraphics[width=\textwidth]{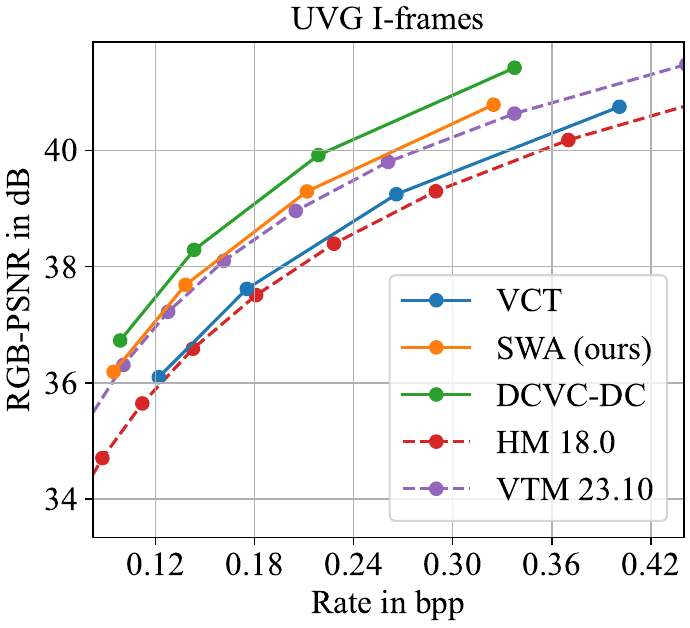}
        \caption{}
    \end{subfigure}
    \begin{subfigure}[b]{0.325\textwidth}
        \centering
        \includegraphics[width=\textwidth]{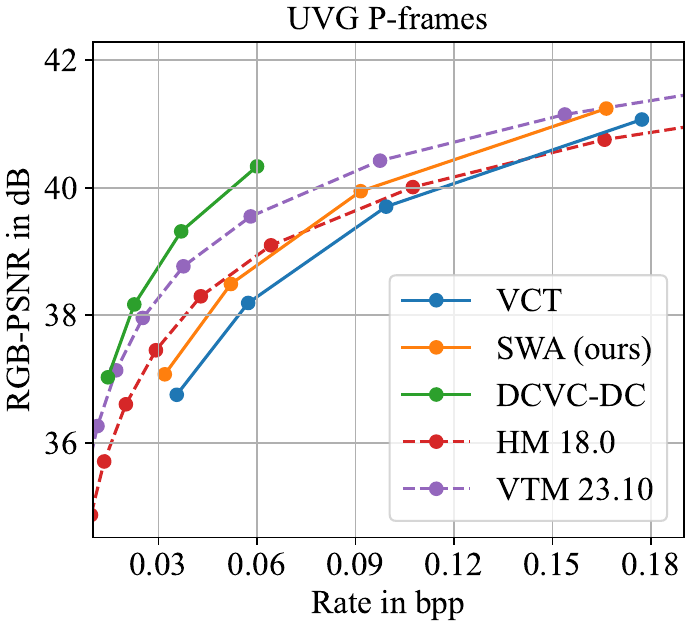}
        \caption{}
    \end{subfigure}
    \begin{subfigure}[b]{0.325\textwidth}
        \centering
        \includegraphics[width=\textwidth]{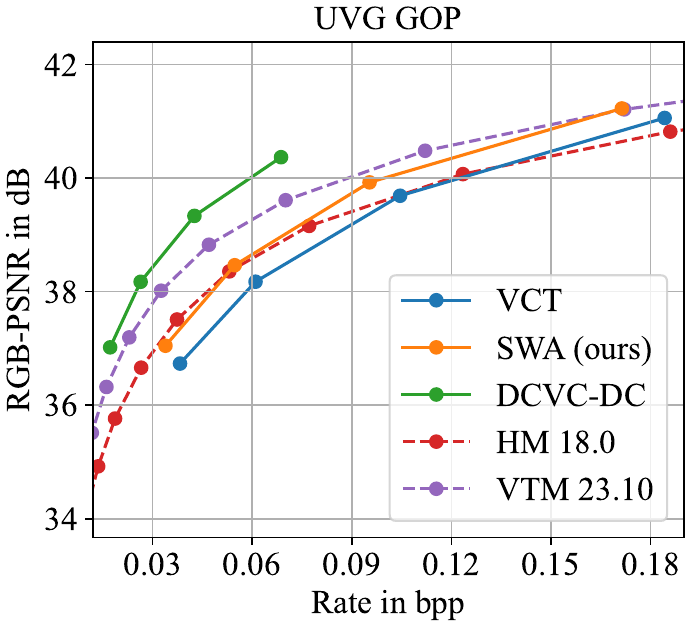}
        \caption{}
    \end{subfigure}
    \captionof{figure}{\gls{RD} curves on the UVG dataset for (a) I-frames, (b) P-frames, and (c) the overall GOP.}
    \label{experiments:results:plots}
\end{figure*}

Our entropy model is a decoder-only transformer that autoregressively predicts the probability distribution for each hyperpixel. Its transformer blocks, shown in Fig. \ref{proposed:sliding:architecture}, are adapted from our \gls{VCT} baseline: the attention mechanism is replaced with \gls{SWA}, and the sinusoidal positional embeddings are removed, as the learned relative biases serve the same function. 
Prediction follows the line-scan order shown in Fig. \ref{proposed:prediction}, and we employ a KV-caching scheme for efficient decoding where each layer caches the key and value tensors from reference frames within its attention window.

The line-scan order creates a challenge at the start of each new row. Following this order, the most recent hyperpixel available at the start of a new row is the last one from the previous row. This position, however, is spatially distant and thus provides poor context for a local attention mechanism. To provide relevant local context, we prepend the hyperpixel from the row directly above to the input sequence, using a zero vector for the first row where no preceding row exists. This ensures context is always drawn from a spatially adjacent neighborhood, which is critical for local attention. This approach differs from \gls{VCT}, which can rely on a single start-of-sequence token per patch.

To support multiple rate points with a single network, we adopt the multi-rate feature transform from DCVC-DC \cite{li_neural_2023}, which produces latents at different scales. To compensate for the scale-invariance of transformers caused by input normalization, we introduce learnable, channel-wise scaling parameters at the transformer's input and its three outputs: the predicted mean and scale of the Gaussian distribution, and an output for the \gls{LRP}. The \gls{LRP}, first introduced by Minnen \textit{et al.} \cite{minnen_channel-wise_2020}, is adopted from the \gls{VCT} baseline and acts as a correction term by predicting the difference between the unquantized and quantized latent.

\section{Experiments}

\subsection{Experimental Setting}

\subsubsection{Models}

For a direct comparison, our \gls{SWA} model and the \gls{VCT} baseline use the same frozen feature transform from the DCVC-DC image model \cite{li_neural_2023}. This isolates performance differences to the entropy model. Our \gls{SWA} model consists of \num{20} transformer layers with a masked \numproduct{5 x 7 x 7} sliding window, providing access to two temporal reference frames in each layer and yielding $\sim$\SI{147}{M} parameters in total. With this configuration, the model closely matches the \gls{VCT} baseline, which has $\sim$\SI{146}{M} parameters and uses \numproduct{8 x 8} patches from two reference frames for its temporal context.

\subsubsection{Training}

Our three-stage training is performed in the RGB color space. First, the feature transform is pre-trained for \num{12} epochs on \numproduct{256 x 256} patches from ImageNet \cite{imagenet-object-localization-challenge}, using a modified entropy model that quantizes latents around zero; the transform's weights are then frozen. Second, the transformer entropy model is trained for \num{12} epochs on \numproduct{256 x 256} clips from OpenVid \cite{nan_openvid-1m_2024}, using every second frame to increase motion. Finally, we fine-tune for two epochs to adapt to larger inputs. To manage complexity, we randomly scale only one dimension per batch (up to \num{40} frames or \num{640} pixels). 
All stages use a batch size of \num{8}. The learning rate follows a cosine anneal from \num{1e-4} to \num{1e-6} for the first two stages, and from \num{2e-5} to \num{1e-6} for fine-tuning.

\subsubsection{Test Conditions}


We evaluate on the UVG \cite{uvg}, MCL-JCV \cite{wang_mcl-jcv_2016}, and HEVC B \cite{Bossen2013common} test sets (\num{96} frames, GOP size \num{32}), following the protocol in \cite{li_neural_2023}. PSNR is calculated in the RGB domain for all comparisons. Source YUV420 videos are converted to RGB for the learned models using bi-linear upsampling and the BT.709 matrix. For comparison, we also test HM 18.0 and VTM 23.10. To enable a direct comparison of P-frame performance against our model, both models are used in their respective low-delay P configurations, which excludes B-frames. They are tested on the intermediate YUV444 data from the upsampling step; for HM 18.0, this requires the RExt profile. The decoded YUV444 output is then converted to RGB for the final PSNR calculation.

\subsection{Results}

\begin{figure}[!t]
    \centering
    \includegraphics[width=0.99\linewidth]{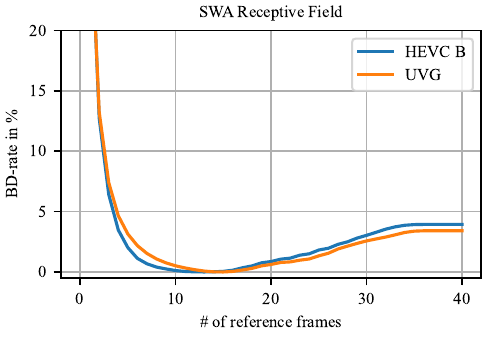}
    \caption{Relative \gls{BD}-rate as a function of the number of reference frames for the HEVC B and UVG datasets. The optimal context size for each sequence is used as the \SI{0}{\percent} reference.}
    \label{experiments:results:receptive}
\end{figure}

As shown by the \gls{RD} curves in Fig. \ref{experiments:results:plots} and the \gls{BD}-rate savings in Table \ref{experiments:results:table}, our proposed \gls{SWA} model significantly outperforms the \gls{VCT} baseline across all test sets. We achieve bitrate reductions of up to \SI{22.6}{\percent} for I-frames and \SI{18.6}{\percent} for P-frames. This includes a PSNR improvement of up to \SI{0.3}{\deci\bel} at lower bitrates, which, given the shared and frozen feature transform, is solely attributable to our entropy model's more effective \gls{LRP}. The overall RD gain stems from two key architectural improvements over \gls{VCT}. First, the local context gathered from our sliding window's uniform receptive field is more effective than the context from the irregular receptive fields of \gls{VCT}'s patches. Second, our unified decoder-only architecture is more efficient; unlike \gls{VCT}'s separated encoder-decoder design, it processes spatial and temporal context jointly and utilizes all layers for spatial modeling. These architectural advantages are particularly pronounced for I-frames, where they are compounded by the fact that our patchless design only lacks context at the very first hyperpixel of the frame, whereas \gls{VCT} faces this issue at the start of every patch. 

While our approach marks a significant improvement over the learned baseline, a performance gap to the state-of-the-art codec, DCVC-DC, remains, particularly for P-frames. This gap is expected, as our model was deliberately kept architecturally simple for a fair comparison to \gls{VCT}. Specifically, this design lacks mechanisms to resolve predictive ambiguity when the context suggests multiple plausible patterns. State-of-the-art codecs typically resolve this ambiguity by using either a hyperprior to provide adaptive side-information, or by updating the context with information from already-decoded feature channels within the same hyperpixel. However, we argue this unified context modeling nonetheless provides a foundation for future codecs to surpass the current state of the art. Unlike methods with separate spatial and temporal modules, its combined context extraction enables a more adaptive context model. By jointly processing all available information, the model can learn how content in the current frame should guide the selection of relevant temporal context, a more flexible approach than the fixed motion estimation of many traditional and learned codecs. Integrating these missing mechanisms is therefore a critical direction for future research.

\subsection{Ablation Study}

To analyze our model's sensitivity to temporal context, we conducted an ablation study on \numproduct{1024 x 1024} center crops from the lower-framerate HEVC B and the high-framerate UVG datasets, shown in Fig. \ref{experiments:results:receptive}. These datasets were chosen to observe the model's behavior on sequences with different temporal characteristics. We varied the number of reference frames from \num{0} to \num{40}, the model's maximum theoretical receptive field, and measured \gls{BD}-rate relative to the optimal context length for each dataset.

The results show that the performance improvement is not monotonic. As expected, performance initially improves dramatically as temporal context is added. However, an optimal context size exists, after which additional reference frames become detrimental. This optimum occurs at \num{13} frames for the lower-framerate HEVC B dataset and \num{15} frames for the higher-framerate UVG dataset, as its consecutive frames are more correlated. Compared to these optimal points, using the full \num{40}-frame context incurs a \gls{BD}-rate penalty of  \SI{3.9}{\percent} and \SI{3.4}{\percent}, respectively. While performance peaks early, we observed that the model remains sensitive to context changes up to approximately \num{35} frames. This confirms that the model's effective receptive field can grow with model depth, but it also suggests the model struggles to disregard distant or irrelevant information. While reducing the temporal kernel size is a potential solution, it presents a trade-off: limiting degradation from irrelevant context at the cost of less efficient long-range information propagation. Although our model's receptive field grows with depth, this sequential accumulation of context may still not be as efficient as information that attention layers can access directly.

\subsection{Complexity}

The theoretical complexities are analyzed in Table \ref{experiments:complexity:overview}. Our \gls{SWA}-based model reduces total decoder kMACs by a factor of \num{2.8} compared to \gls{VCT}, with the entropy model itself becoming nearly \num{3.5} times more efficient. This gain stems from our patchless design, which processes each reference hyperpixel only once, unlike \gls{VCT}'s overlapping window scheme where each is processed four times. However, a practical limitation of our model is its sequential decoding process, a consequence of the single line-scan prediction across the entire frame. In contrast, while \gls{VCT} also uses a sequential line-scan within each patch, its independent patches can be decoded in parallel. Therefore, adopting a parallel decoding scheme is the critical next step to translate the theoretical efficiency of this architecture into higher decoding speed. This can be implemented by modifying the kernel's causal mask to enable parallel spatial prediction, using a prediction pattern similar to \cite{li_neural_2023}. 

A key architectural difference to the motion-based DCVC-DC is the relative encoder complexity. As shown in Table \ref{experiments:complexity:overview}, DCVC-DC's encoder is more complex than its decoder because it must also run the feature synthesis transform to generate pixel-domain reference frames for motion estimation. In contrast, our model and \gls{VCT} operate directly on reference latents, resulting in less complex encoders.

\begin{table}[t]
    \centering
    \caption{Encoder and decoder complexity comparison in kilo-MACs per pixel (kMACs/px). For DCVC-DC, the complexity is averaged over a \num{32}-frame GOP.}
    \label{experiments:complexity:overview}
    \sisetup{
        table-format = 4.2, 
        detect-weight,
        mode = text
    }
    \begin{tabular}{l l S S}
        \toprule
        \textbf{Model} & \textbf{Component} & {\textbf{Encoder}} & {\textbf{Decoder}} \\
        & & {\textbf{kMACs/px}} & {\textbf{kMACs/px}} \\
        \cmidrule(lr){3-4}
        \textbf{VCT}     & Feature Transform & 154.08 & 240.34 \\
                         & Entropy Model     & 2080.30 & 2080.30 \\
                         \cmidrule(l){2-4}
                         & \textbf{Total}    & \bfseries 2234.38 & \bfseries 2320.64 \\
        \midrule
        \textbf{SWA (ours)}     & Feature Transform & 154.08 & 240.34 \\
                         & Entropy Model     & 598.46 & 598.46 \\
                         \cmidrule(l){2-4}
                         & \textbf{Total}    & \bfseries 752.54 & \bfseries 838.80 \\
        \midrule
        \textbf{DCVC-DC} & \textbf{Total (GOP Avg.)} & \bfseries 1307.61 & \bfseries 901.53 \\
        \bottomrule
    \end{tabular}
\end{table}

\section{Conclusion and Outlook}


This work introduces 3D \gls{SWA} to address the inefficiency of patch-based transformers like \gls{VCT}. Our patchless, fully autoregressive entropy model eliminates the redundant computations of \gls{VCT}'s overlapping window scheme, establishing a more performant and efficient foundation for transformer-based video codecs. This new approach achieves \gls{BD}-rate savings of up to \SI{18.6}{\percent} against the \gls{VCT} baseline and reduces theoretical decoder complexity by a factor of \num{2.8}, with the entropy model itself becoming nearly \num{3.5} times more efficient.


Despite these improvements, a performance gap to state-of-the-art learned codecs remains, as our model was deliberately kept architecturally simple for a fair comparison. Our analysis, however, also points to clear paths for future enhancement. Future work should focus on adapting \gls{SWA} for a parallel prediction scheme to improve decoding speed, investigating the optimal temporal context length, and integrating additional modules, such as a hyperprior, to leverage the architecture's inherent flexibility and ultimately surpass the current state of the art.

\bibliographystyle{IEEEtran}
\bibliography{bib.bib}

\end{document}